\documentclass[MSNbibl,number,dvips]{arxstspdf}
\usepackage{flushend}
\usepackage{stfloats}

\volume{26}
\issue{4}
\pubyear{2011}
\firstpage{598}
\lastpage{600}
\doi{10.1214/11-STS356A}
\referstodoi{10.1214/11-STS356}

\makeatletter

\newtheorem{theorem}{Theorem}[section]
\newtheorem{lemma}{Lemma}[section]

\newproclaim{definition}{Definition}[section]

\makeatother

\begin{document}
\begin{frontmatter}
\vspace*{6pt}
\title{Discussion of ``Multiple Testing for Exploratory Research'' by J. J. Goeman and A. Solari}
\runtitle{Discussion}

\begin{aug}
\author[a]{\fnms{Ruth} \snm{Heller}\corref{}\ead[label=e1]{ruheller@post.tau.ac.il}}
\runauthor{R. Heller}

\affiliation{Tel-Aviv university}

\address[a]{Ruth Heller is Senior Lecturer, Department of Statistics and Operations
Research, Tel-Aviv University, Tel-Aviv, Israel\\ \printead{e1}.}

\end{aug}

%
\begin{abstract}
Goeman and Solari [\textit{Statist. Sci.} \textbf{26} (2011) 584--597] have
addressed the interesting topic of multiple testing for exploratory research,
and provided us with nice suggestions for exploratory analysis. They
defined properties that an inferential procedure
should have for exploratory analysis: the procedure should be {\em
mild\textup{,} flexible} and {\em post hoc}.
Their inferential procedure gives a lower bound on the number of false
hypotheses among the selected hypotheses,
and moreover whenever possible identifies elementary hypotheses that
are false. The need to estimate a lower bound
on the number of false hypotheses arises in various applications, and
the partial conjunction approach was
developed for this purpose in \textit{Biometrics} \textbf{64} (2008)
1215--1222 (see also
\textit{Philos. Trans. R. Soc. Lond. Ser. A} \textbf{367} (2009)
4255--4271 for more details). For example, in a
combined analysis of several studies that examine the same problem, it
is of interest to give a lower bound
on the number of studies in which the finding was reproduced. I will
first address the relation between the
method of Goeman and Solari and the partial conjunction approach. Then
I will discuss possible extensions
and address the issue of exploration in more general settings, where
the local test may not be defined
in advance or where the candidate hypotheses may not be known to begin with.

\end{abstract}


\end{frontmatter}

\section{Relation to the Testing of Partial Conjunction
Hypotheses}\label{sec-conj}

Let $H_1,\ldots, H_n$ be the elementary hypotheses. The idea of giving
a lower bound on the number of false elementary hypotheses (or
equivalently an upper\break bound on the number of true elementary
hypotheses) appears in \cite{conj}, and is closely related to the tests
of partial conjunction hypotheses.
The partial conjunction null hypothesis\vadjust{\goodbreak} $H^{u/n}$ in \cite{conj} asks
whether fewer than $u$ of the elementary hypotheses are false, and the
alternative hypothesis is that at least $u$ of the elementary
hypotheses are false. Testing whether $H^{u/n}$ is false at a
significance level $\alpha$ in order (i.e., for $u=1,2,\ldots$) results
in a $1-\alpha$ confidence lower bound on the number of false
elementary hypotheses:

\begin{theorem}\label{theorem1}
Let $p^{u/n}$ be a partial conjunction p-value for
testing $H^{u/n}$. Let $u_{\max} = \max\{u\dvtx
p^{i/n}\leq\break\alpha\ \forall i=1,\ldots,u \}$. Then
with $1-\alpha$ confidence, the true number of false hypotheses is
in $[u_{\max},n]$.
\end{theorem}

\begin{pf}
Let $k$ be the true number of false elementary hypotheses. If $k=n$,
that is, all
elementary hypotheses are false, there is nothing to prove. If $k<n$,
\begin{eqnarray*}
\operatorname{Pr}(k\geq u_{\max}) &=& 1-\operatorname{Pr}(k<u_{\max})
\\
&=& 1-\operatorname{Pr}\bigl(P^{(k+1)/n}\leq\alpha\bigr)
\geq1-\alpha.\quad~\qed%
\end{eqnarray*}%
\noqed\end{pf}

The lower bound $u_{\max}$ above is identical to the lower bound of
Goeman and Solari (denoted by\break $f_\alpha\{1,\ldots,n \}$ in their
paper), when the full set of elementary hypotheses is considered.
Moreover, the shortcuts suggested by Goeman and Solari are equivalent
to the tests of partial conjunction hypotheses suggested in \cite
{conj}, that do not require examination of all $\bigl( {{n}\atop{n-u+1}}\bigr)$
intersection hypotheses for the test of~$H^{u/n}$ , but rather require
only testing the subset of $n-u+1$ intersection hypotheses that
correspond to the $n-u+1$ least significant elementary hypotheses
$p$-values. Specifics follow.

Reference \cite{conj} suggested methods for combining the $p$-values for
testing $H^{u/n}$ that are based on {\em sufficient combining
functions}.
\begin{definition} $f(U_1,\ldots,U_{m})$ is a sufficient\break
combining function from $ \Re^{m} \rightarrow\Re$ if it has the
following properties:
\begin{enumerate}
\item If $U_i'\!\geq\!U_i$, then $f(U_1,\ldots, U_{i-1},U_i',U_{i+1},\ldots
, U_{m})\!\geq\!f(U_1,\ldots, U_{i-1},U_i,U_{i+1},\ldots, U_{m})$, that
is, $f$ is an increasing function of its components.
\item If $U_i$ is uniformly distributed or stochastically larger than
the uniform, that is, $U_i\mathop{\succeq}\limits_
{\mathrm{st}} U(0,1)\ \forall i=1,\ldots, n$, then $f(U_1,\ldots,
U_{m})\mathop{\succeq}\limits_
{\mathrm{st}} U(0,1)$.
\end{enumerate}
\end{definition}


Let $p_{(1)}\leq\cdots\leq p_{(n)}$ be the sorted $p$-values.
The following lemma gives the guiding principle for the $p$-values
suggested in \cite{conj} for testing the
partial conjunction hypothesis:
\begin{lemma}\label{thmConj}
Let $f(U_1,\ldots,U_{n-u+1})$ be a sufficient combining function
from $ \Re^{n-u+1} \rightarrow\Re$. Let $p^{u/n}$ be the result of
combining the largest $n-u+1$ $p$-values using the function $f$,
that is, $p^{u/n} = f(p_{(u)},\ldots,\break p_{(n)})$. Then $\operatorname{Pr}(P^{u/n}\leq
\alpha)\leq\alpha$ if $H^{u/n}$ is true.
\end{lemma}

For example, if the $p$-values are independent the $p$-value motivated
by the Fisher method for testing~$H^{u/n}$ is
\[
p^{u/n} = \operatorname{Pr}\Biggl(\chi^2_{2(n-u+1)}\geq-2\sum_{i=u}^{n} \log p_{(i)}\Biggr).
\]
Finding $u_{\max}$ using the partial conjunction test $p$-values based
on Fisher's method will give the same result as the procedure in
Section 4.1 of Goeman and Solari, when the full set of elementary
hypotheses is considered.

Similarly, if a set $R\subset\{1,\ldots,n\}$ is selected a priori,
then the lower bound on the number of false hypotheses may be found by
testing in order the partial conjunction hypotheses $p^{u/|R|},
u=1,2,\ldots,$ where $|R|$ is the cardinality of $R$. If the set $R$ is
selected post hoc, then the lower $1-\alpha$ confidence bound on the
number of false hypotheses may be lower than the bound resulting from
the above procedure because of the selection effect, and the procedures
suggested by Goeman and Solari can be used to adjust for the selection effect.


\section{Multiple Families of Hypotheses in~Exploratory
Research}\label{sec2}

In \cite{conj}, the partial conjunction approach was used to estimate
the lower bound on the number of false hypotheses when a large number
of such lower bounds need to be estimated simultaneously. In multiple
testing for exploratory research, a similar problem may arise.
Consider, for example, a large genomics study, where the signal in many
genes (or SNPs) are measured simultaneously. In order to select genes
(or SNPs) for follow-up, the researcher may want to select a subset of
promising genes from prespecified regions in the genome. In such a
problem, in each region a subset of promising genes (or SNPs) may be
selected by exploration of that region.

When exploring multiple families of hypotheses, in order to limit the
total number of false leads, the decision about the subset of
hypotheses selected for follow-up in each family may be affected by the
estimated lower bounds on the number of false null hypotheses in the
subsets selected in other families of hypotheses. Moreover, the
researcher may be interested in a lower bound on the number of false
leads at the level of families rather than at the level of elementary
null hypotheses. These are natural extensions to the problem addressed
by Goeman and Solari, where multiple testing may be applied to multiple
families of hypotheses in an exploratory manner.

\section{The Choice of the Local Test}\label{sec3}
The approach of Goeman and Solari assumes that the test of each
intersection hypothesis is known in advance. However, it may be
difficult to decide which local test is best without first looking at
the data.

In some applications, we may not always have a~good statistic in mind
for evaluating an elementary null hypothesis.
We may need to explore the data in order to decide on a good test
statistic for testing the null hypothesis. However, when testing the
elementary hypothesis on the data explored to decide on the test, the
test is no longer a valid test in the sense that there is no guarantee
it preserves the level of the test.

Moreover, when we have several elementary hypotheses of interest and we
want to test their intersection hypothesis, how should the test
statistic be chosen? Different tests will have power against different
alternatives. Even if we limit ourselves to tests that are based on
combining functions of the elementary hypotheses $p$-values, different
functions are better capable of detecting different patterns of
evidence against the intersection null hypothesis, and the differences
among them can be large (see, e.g., \cite{loughin} and \cite{donoho}). Because
no single combining function can be best under all circumstances, in
exploratory analysis the researcher may choose a combining function by exploring
different combining methods. The chosen method may then be used on
data from follow-up studies. However, for testing the intersection
hypotheses on the data explored, the test is no longer a valid test.

Therefore, if the data are explored to select which local test to use,
the confidence sets may no longer have the correct level and may be
misleading. Nevertheless, the use of multiple testing for selecting
hypotheses for follow-up is still valuable as a tool, even though it is
not possible to quantify the number of false leads in the selected
subset of hypotheses for follow-up.

\section{The Practice of Exploratory Research}\label{sec4}
Even when multiple comparisons issues are addressed, still studies are
too often
not reproducible (see \cite{Ioannidis05}) and scientists follow too
many false leads. This may be because together with advances
in multiple comparisons over the years, there have been many advances
in how
data can be explored. The multiple comparisons correction is possibly
done only
on a subset of hypotheses without intention. From sophisticated (and
even simple)
graphical displays, a hypothesis may be generated. But how can one
quantify then how
many potential hypotheses have actually been tested before selecting
the particularly
interesting one based on the picture?
If the user cannot quantify how many hypotheses may be looked at in the
exploratory
stage, how should the data be analyzed to select promising hypotheses
to follow up on while still quantifying the error in terms of a~lower
bound on the number of false null hypotheses?\looseness=-1


One possibility is to define the hypotheses on part of the data by
creative exploratory analysis and then apply the multiple testing
procedure on the rest of the data (see \cite{Cox75}). The problem is
that by testing only part of the data we lose power. Therefore,
a~modest change in current practice may be the following: to set aside
only the amount of data that the investigator is willing to spare for
the purpose of generation of hypotheses and in order to decide what
local test to use for each hypothesis. So, for example, from a study of
500 subjects the investigator may be willing to set aside 100 subjects,
and from a~sample size of 100 perhaps only 15 subjects may be set aside
for hypothesis generations. Once the hypotheses and tests of hypotheses
have been decided upon, the procedure of Goeman and Solari may be
applied. This process is {\em mild, flexible} and {\em post hoc}
without losing all ability to quantify the confidence on the estimated
number of false positives among the selected hypotheses.

\section*{Acknowledgments}
Supported by the Israel Science Foundation (ISF) Grant no. 2012896.


\end{document}